\documentclass[aps,pre,floats,twocolumn,showpacs]{revtex4-1}
\usepackage[dvips]{graphicx}
\usepackage{epsfig}
\usepackage{subfigure}
\usepackage{amsmath}
\usepackage{epsfig}
\usepackage{color}\newcommand{\ID}{\mathbb{1}}
\usepackage{easybmat}
\usepackage{multirow,bigdelim}
\usepackage{bbold}

\newcommand{\bsigma}{\mbox{\protect\boldmath $\sigma$}}

\newcommand{\bss}{\mbox{\protect\boldmath $s$}}
\newcommand{\calL}{{\cal L}}
\newcommand{\calG}{{\cal G}}
\newcommand{\Trace}{{\rm Tr}}

 \newcommand{\sgn}{\mathop{\mathrm{sgn}}}
\begin{document}

\title{Learning and inference in a nonequilibrium Ising model with hidden nodes}

\author{Benjamin Dunn}
\email{benjamin.dunn@ntnu.no}
\affiliation{The Kavli Institue for Systems Neuroscience, NTNU, 7030 Trondheim}
\author{Yasser Roudi}
\email{yasser.roudi@ntnu.no}
\affiliation{The Kavli Institue for Systems Neuroscience, NTNU, 7030 Trondheim}
\affiliation{NORDITA, KTH Royal Institute of Technology and Stockholm University, 10691 Stockholm, Sweden}

\begin{abstract}
We study inference and reconstruction of couplings in a partially observed
kinetic Ising model. With hidden spins, calculating the likelihood of a sequence of
observed spin configurations requires performing a trace over the configurations of the
hidden ones. This, as we show, can be represented as a path integral. Using this 
representation, we demonstrate that systematic approximate inference and learning rules can be derived using
dynamical mean-field theory. Although naive mean-field theory leads to an unstable learning rule, 
taking into account Gaussian corrections allows learning the couplings involving hidden nodes. It also improves
learning of the couplings between the observed nodes compared to when hidden nodes are ignored.
\end{abstract}
\maketitle

\section{Introduction}
Within the statistical mechanics community, a
significant body of recent work deals with parameter learning in statistical models.
This work typically focuses on the
canonical setting of the inverse Ising problem: given a set of configurations from an
equilibrium
\cite{Kappen98,*Tanaka98,*Roudi09-3,*Cocco2011adaptive,*Ricci2012,*Nguyen2012,*Nguyen2012PRL}, or
nonequilibrium \cite{Roudi11,*Mezard11,*Zeng11,*zhang2012inference} Ising model, how can we
find the interactions between the spins? These inverse problems can be difficult to solve
using brute force approaches \cite{Ackley85} and thus for practical purposes, 
the results of such theoretical works offering approximate learning and inference methods
are of great importance. The recent interest in using statistical models to study
high-throughput data from neural \cite{Schneidman06,Shlens06,Cocco09}, genetic \cite{Lezonetal06} and
financial \cite{bury2012markets} networks further emphasizes the necessity of 
developing efficient approximate inference methods.

Despite its practical relevance, however, little
has been done on learning and inference in models with hidden spins using statistical
physics. Learning the parameters of a statistical model in the presence of
hidden nodes and inferring the state of hidden variables lie at the heart of difficult
problems in statistics and machine learning. It is well known that for statistical modeling of data, hidden variables most
often improve model quality \cite{bishop1998latent,Pearl00}. Furthermore, when used for network
reconstruction and graph identification, ignoring hidden nodes may cause the model to
incorrectly identify/remove couplings between observed nodes when such couplings do not
exist/actually exist. For these reasons, it is quite important to study models with 
hidden nodes, both from a theoretical perspective and for
applications to real data where usually only a fraction of the 
system is observeable \cite{Schneidman06,Shlens06,Lezonetal06}.

In this paper, we study the reconstruction of the couplings in a nonequilibrium Ising model
with hidden spins and the inference of the state of these spins. Theoretically, the couplings can
be reconstructed by calculating and maximizing the likelihood of the data which involves
performing a trace over the configurations of the hidden spins at all times: an impossible
task even for moderately sized data sets and networks. This is, in fact, a general problem,
not only for a nonequilibrium Ising model, but for any kinetic model with hidden variables.
It is, however, natural to use a path integral approach for doing this trace, and this is
what we do here. We show that efficient approximate learning and inference can be
developed by evaluating the log-likelihood in this way and employing mean-field
type approximations. 

In what follows, after formulating the problem, we first evaluate the likelihood of the data using the saddle point approximation 
to the path integral. We find that the result allows inference of the state of hidden 
variables if the couplings are known but cannot be used for learning the couplings. We then
perform a Gaussian correction to this saddle point approximation to derive 
Thouless-Anderson-Palmer (TAP) like equations \cite{Thouless77} for our problem. Our numerical 
results show that successul learning of the couplings, even those that involve hidden nodes,
can be achieved using the corrected learning rules. We find that 
with enough data and using these equations even hidden-to-hidden connections can be reconstructed. 

Although in this paper our focus is on the paradigmatic case of the Ising model, the approach taken
here is quite general and can be adapted to many other popular statistical models.

\section{Formulation of the problem}
Consider a network of binary spin variables subject to synchronous
update dynamics. Suppose some of these nodes are observed for $T$ time steps,
while the other nodes are hidden. We denote the state of the observed spins at time $t$ by
$\bss(t)=\{s_i(t)\}$ and the hidden ones by $\bsigma=\{\sigma_a(t)\}$. For clarity, we always use
the subscripts $i,j, \dots$  for the observed and $a,b, \dots$ for the hidden ones. We 
consider the following transition probability
\begin{subequations}
\begin{eqnarray}
&&p[\{\bss,\bsigma \}(t+1)|\{\bss,\bsigma \}(t)]=\\
&&\hspace{30pt} \exp[\sum \nolimits_i s_i(t+1)g_i(t)+\sum \nolimits_a \sigma_a(t+1) g_a(t)] Z(t)^{-1} \nonumber\\
&&Z(t)=\prod_{i,a} 2 \cosh[g_i(t)]\ 2 \cosh[g_a(t)]
\end{eqnarray}
\label{DynSK}
\end{subequations}
where $g_i(t)$ and $g_a(t)$ are the fields acting on observed spin $i$ and hidden spin $a$ at time $t$, respectively
\begin{subequations}
\begin{eqnarray}
&&g_i(t)=\sum \nolimits_j J_{ij} s_j(t)+\sum \nolimits_b J_{ib} \sigma_b(t)\label{fieldi}\\
&&g_a(t)=\sum \nolimits_j J_{aj} s_j(t)+\sum \nolimits_b J_{ab} \sigma_b(t)\label{fielda},
\end{eqnarray}
\label{fields}
\end{subequations}
$J_{ij},J_{ai},J_{ia}$ and $J_{ab}$ are the observed-to-observed, observed-to-hidden,
hidden-to-observed and hidden-to-hidden connections. These couplings need not be symmetric and, thus,
the network may never reach equilibrium. Although, here
we consider the case of zero external field, our derivations follow also for nonzero fields. 

Given this model, we would like to answer the following questions:
What can we say about the state of hidden spins from the observations? How can we learn
the various couplings in the network?

Answering these questions requires finding the likelihood of the data.
Optimizing the likelihood with respect to $J$s yields the maximum likelihood value of the couplings.
The posterior over the state of hidden nodes given the observed nodes is 
\begin{equation}
p[\{\bsigma(t)\}^T_{t=1}|\{\bss(t)\}^T_{t=1}]=\frac{p[\{\bsigma(t)\}^T_{t=1}\ ,\{\bss(t)\}^T_{t=1}]}{p[\{\bss(t)\}^T_{t=1}]}\nonumber
\end{equation}
which also requires calculating the likelihood. 

The likelihood of the observed spin configurations under the model Eq.\ \ref{DynSK} is
\begin{equation} 
p[\{\bss(t)\}^T_{t=1}]=\Trace_{\sigma}\prod_t  p[\{\bss,\bsigma \}(t+1)|\{\bss,\bsigma \}(t)]
\end{equation}
and involves a trace over $\{\bsigma(1),\dots,\bsigma(T) \}$ which is difficult to do. 

To perform this trace analytically, we consider the following functional
\begin{equation} 
\calL[\psi]\equiv \log \Trace_{\sigma} \prod_t e^{\sum_{a} \psi_a(t)\sigma_a(t)} p[\{\bss,\bsigma \}(t+1)|\{\bss,\bsigma \}(t)].
\label{LL}
\end{equation}
which, using the definition of the transition probability from Eq.\ \ref{DynSK}, can be written as
\begin{eqnarray}
&&\hspace{-0.5cm}  \calL[\psi]= \log \Trace_{\sigma} \exp\left[Q[\bss,\bsigma]+\sum_{a,t} \psi_a(t)\sigma_a(t)\right]\label{LQ}\\
&&\hspace{-0.5cm} Q= \sum_{i,t} s_i(t+1) g_i(t)+\sum_{a,t} \sigma_a(t+1) g_a(t) \nonumber\\
&&\hspace{1cm} -\sum_{i,t} \log 2\cosh[g_i(t)]-\sum_{a,t}\log 2\cosh[g_a(t)].\nonumber
\end{eqnarray}
The log-likelihood of the data is recovered at $\psi\to 0$. Furthermore, $\calL[\psi]$ acts as the generating 
functional for the posterior over the hidden spins: $m_a(t)$, 
the mean magnetization of the hidden spin $a$ at time $t$ given the data, can be found using 
\begin{equation} 
m_a(t)=\lim_{\psi\to 0} \mu_a(\psi,t), \ \ \ \  \mu_a(\psi,t)=\frac{\partial \calL[\psi]}{\partial \psi_a(t)}.
\label{ma}
\end{equation}
Higher-order derivatives of $\calL$ with respect to $\psi$ yield higher-order correlation functions of the hidden spins.

To develop mean-field approximations, we first perform the trace in Eq.\ \ref{LQ}
using the standard approach for dealing with a nonlinear action in the local fields \cite{Coolen2001619,OpperWinther2001}: we
write the term inside the log in Eq.\ \ref{LQ} as a path integral over $g_i(t)$ and
$g_a(t)$ enforcing their definitions in Eq.\ \ref{fields}
by inserting $\delta[g_i(t)-\sum_j J_{ij} s_j(t)-\sum_b J_{ib} \sigma_b(t)]$
and $\delta[g_a(t)-\sum_j J_{aj} s_j(t)-\sum_b J_{ab} \sigma_b(t)]$ in the integral. Using the 
integral representation of the delta functions, $\calL$ can be written as
\begin{subequations}
\begin{eqnarray}
&&\calL[\psi]= \log  \int D\calG \exp[\Phi] \label{likelihoodint}\\ 
&&\Phi=\log \Trace_{\bsigma} \exp[Q+\Delta+\sum_{a,t} \psi_a(t) \sigma_a(t)]
\end{eqnarray}
\label{likelihood2}
\end{subequations}
where $\calG=\{g_i,\hat{g}_i,g_a,\hat{g}_a\}$. $\Delta$, as well as $\hat{g}_i$ and $\hat{g}_a$,
come from the integral representation of the $\delta(\cdot)$ functions used to enforce
the definitions in Eq.\ \ref{fields}
\begin{eqnarray}
&&\hspace{-0.5cm}\Delta=\sum_{i,t}i\hat{g}_i(t)[g_i(t)-\sum_j J_{ij} s_j(t)-\sum_b J_{ib} \sigma_b(t)]\\
&&\ \ \ \ \ +\sum_{a,t} i\hat{g}_a(t)[g_a(t)-\sum_j J_{aj} s_j(t)-\sum_b J_{ab} \sigma_b(t)]\nonumber
\end{eqnarray}
The crucial thing about this rewriting of $\calL$ is that the resulting action $\Phi$ is
now linear in $\sigma$ and we can thus easily perform the trace over $\sigma$. The
ability to perform the trace here comes at the cost of the high-dimensional integral
over $\calG$ in Eq.\ \ref{likelihoodint}. The integral, however, can be 
approximated using the saddle point approximation and the corrections to the saddle point.

\section{Saddle-point approximation} 
The saddle point 
values of $g_a,g_i,\hat{g}_a$ and $\hat{g}_i$ can be derived by putting the derivatives
of $\Phi$ with respect to these variables to zero. For $\psi=0$, the saddle point equations
read
\begin{subequations}
\begin{eqnarray}
&&\hspace{-0.8cm}i\hat{g}_i(t)=\tanh[g_i(t)]-s_i(t+1)\label{ghsaddle1}\\
&&\hspace{-0.8cm}i\hat{g}_a(t)=\tanh[g_a(t)]-m_a(t+1)\label{ghsaddle2}\\
&&\hspace{-0.8cm} g_i(t)=\sum \nolimits_j J_{ij} s_j(t)+\sum  \nolimits_b J_{ib}m_b(t) \label{gisaddle}\\
&&\hspace{-0.8cm} g_a(t)=\sum  \nolimits_j J_{aj} s_j(t)+\sum  \nolimits_b J_{ab}m_b(t)\label{gasaddle}
\end{eqnarray}
\label{gsaddles}
\end{subequations}
in which $m_a(t)$ satisfy the self-consistent equations
\begin{eqnarray}
&&\hspace{-10pt}m_a(t)=\label{selfconst}\\
&&\tanh\big[g_a(t-1)-i\sum \nolimits_j \hat{g}_j(t) J_{ja}-i\sum \nolimits_b \hat{g}_b(t) J_{ba}\big]\nonumber
\end{eqnarray}
Let us take a moment to describe the physical meaning of these equations.
The saddle point equations for $g_i$ and $g_a$, Eq.\ \ref{gisaddle} and \ref{gasaddle}, are just
the mean values of the local fields at the mean magnetizations $m_a(t)$ of the hidden
variables (see Eq.\ \ref{fields}). Examining Eq.\ \ref{ghsaddle1} and 
\ref{ghsaddle2}, one can identify $-i\hat{g}_i(t)$ and $-i\hat{g}_a(t)$ as errors back
propagated in time. The self-consistent equations for $m_a$, Eq.\ \ref{selfconst}, 
take the form of the usual naive mean-field equations except that back propagated
errors are taken into account.

Given a sequence of configurations from the observed spins,
solving Eq.\ \ref{selfconst} yields the mean magnetization of
the hidden ones given this data. The derivatives of $\Phi$ with respect to the $J$s, 
evaluated at the saddle point solutions, give 
the maximum likelihood learning rules for the couplings within this
saddle point approximation.

\section{Gaussian corrections}
The saddle-point equations and learning rules can be improved by Gaussian corrections,
that is by evaluating the contributions to the path integral from the Gaussian fluctuations 
around the saddle \cite{negele1998quantum,Kholodenko1990}. This will lead
to what can be called as TAP equations for our problem. 
In what follows, by doing this on the Legendre transform of $\calL$,
\begin{equation}
\Gamma[\mu]=\calL-\sum_{a,t} \psi_a(t) \mu_a(t) \label{LG1}
\end{equation}
and using the equation of state,
\begin{equation}
-\psi_a(t)=\frac{\partial \Gamma[\mu]}{\partial \mu_a(t)} \label{stateq}
\end{equation}
we correct the naive mean-field equations for our problem with the hidden spins.

Denoting the saddle point value of $\calL$ by $\calL_0$, 
and performing a Legendre transform we have (see Appendix \ref{appendixI})
\begin{eqnarray}
&&\Gamma_0[\mu]=\sum_{i,t} [s_i(t+1) g_i(t)-\log 2\cosh(g_i(t))]\label{Gamma0}\\
&&\ \ \ \ +\sum_{a,t} [\mu_a(t+1) g_a(t)-\log 2\cosh(g_a(t))]+\sum_{a,t} S[\mu_a(t)]\nonumber
\end{eqnarray}
where $S[\mu_a(t)]$ is the entropy of a free spin with magnetization $\mu_a(t)$, and  
$g_i$ and $g_a$ are as in Eq.\ \ref{fields} with $\sigma_b(t)$ replaced by
$\mu_b(t)$. Using Eq.\ \ref{stateq} and putting $\psi=0$, we recover Eq.\ \ref{selfconst},
as we should.  

The Gaussian corrections can be performed as follows; for details
see Appendix \ref{appendixII}. We first perform the Gaussian integral around
the saddle point of $\calL$. This modifies $\calL_0$ to 
\begin{equation}
\calL_1=\calL_0-\frac{1}{2}\log \det[\partial^2 \calL] 
\end{equation}
where $\partial^2 \calL$ is the Hessian, containing the second
derivatives of $\calL$ with respect to $g$ and $\hat{g}$,
calculated at their saddle point values. Keeping terms quadratic in $J$ we then derive the 
modified equations for $m_a(t)$ by first calculating $\partial \calL_1/\partial \psi_a(t)$.
Expressing $\calL_1$ in terms of the new $m_a(t)$, again keeping terms
quadratic in $J$, and finally performing the Legendre transform, we get
\begin{eqnarray}
&&\Gamma_1[m]=\Gamma_0[m]-\frac{1}{2}\sum_{i,a,t} [1-\tanh^2 g_i(t)] J^2_{ib}[1-m^2_b(t)]\nonumber\\
&&\hspace{0.4cm}-\frac{1}{2}\sum_{a,b,t} [m^2_a(t+1)-\tanh^2 g_a(t)] J^2_{ab}[1-m^2_b(t)].\label{GammaTAP}
\end{eqnarray}
where now, $g_i$ and $g_a$ are as in Eq.\ \ref{fields} but with $m_a$ instead of $\sigma_a$.  
Eq.\ \ref{GammaTAP} together with the equation of state at $\psi \to 0$,
\begin{equation}
\frac{\partial \Gamma_1[m]}{\partial m_a(t)}=0,
\label{maTAP}
\end{equation}
yields the new self-consistent equation for $m_a(t)$. The couplings can then be found 
by optimizing $\Gamma_1$ with respect to the couplings. 

In sum, the corrected TAP-learning will
be the following EM like iterative algorithm:
at each step, solve the equations for $m_a$ derived from Eq.\ \ref{maTAP} 
given the current value of the couplings, and then adjust the 
couplings proportional to $\partial \Gamma_1/\partial J$.  

\section{Numerical results}
\begin{figure}[t]
\subfigure{\includegraphics[height=1.6in, width=1.6in]{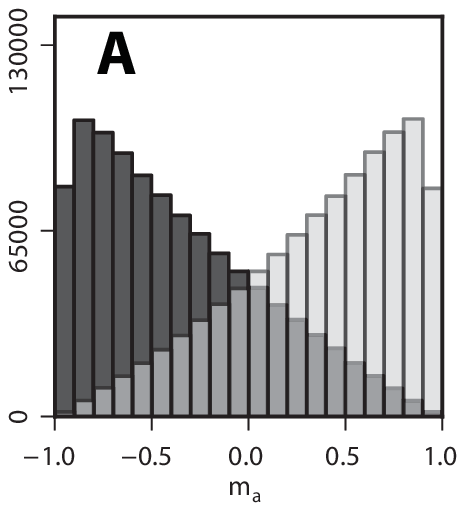}}
\subfigure{\includegraphics[height=1.6in, width=1.6in]{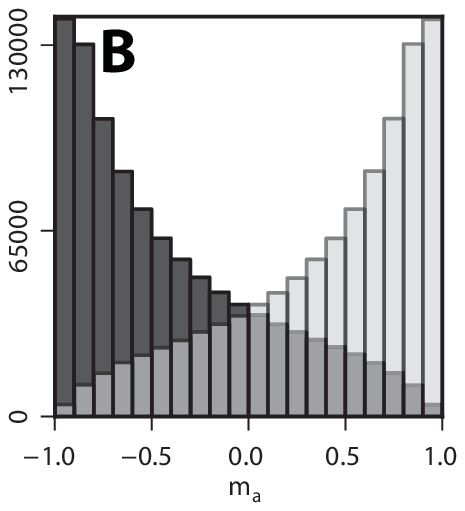}}\\
\subfigure{\includegraphics[height=1.6in, width=2in]{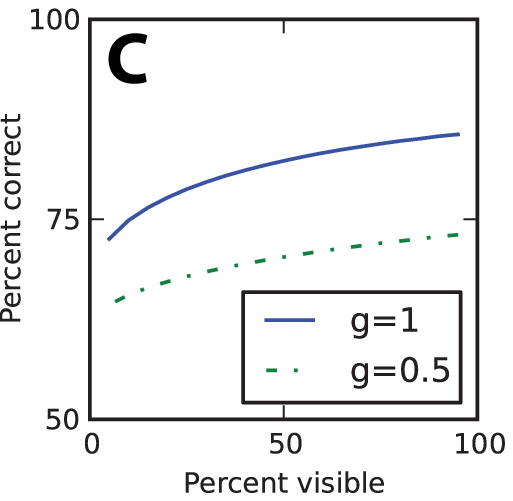}}
\caption{(Color online) Solutions for $m_a$. (A) and (B) show histograms of $m_a(t)$ found from solving
saddle point and TAP equations, Eqs.\ \ref{selfconst}, \ref{maTAP}, given
the correct couplings. Here we had $J_1=0.7$, $20\%$ hidden spins and a data length of $T=10^5$.
The dark gray bars show the histogram of the values for which $\sigma_a(t)$ was actually $+1$ while the
transparent ones show when it was actually $-1$. (C)
shows the percent correct if we use $\hat{\sigma}_a(t)=\sgn(m_a(t))$ to infer the
state of $\sigma_a(t)$ vs. the percentage of the visible part. TAP and
saddle point results were virtually indistinguishable.}
\label{Fig1}
\end{figure}

 We evaluated the saddle point and TAP equations on data
generated from a network of $N=100$ spins with the dynamics in Eq.\ \ref{DynSK} and with
the couplings drawn independently from a Gaussian distribution with variance $J_1^2/N$. We first studied
how much the saddle point Eq.\ \ref{selfconst} and TAP equations Eq.\ \ref{maTAP} tell us
about the state of hidden spins if we know all the couplings. Fig.\ \ref{Fig1} A and B
show the distributions of $m_a(t)$ when $20\%$ of the network is hidden. The solutions to
the TAP equations are more pushed towards $+1$ and $-1$ compared to the saddle
point. Using a simple estimator $\hat{\sigma}_a(t)=\sgn(m_a(t))$ to infer the state of
hidden spin $a$ and time $t$, Fig.\ \ref{Fig1}C shows that the two approximations 
perform equally well in inferring the state of hidden spins. 
\begin{figure}[t]
\subfigure{\includegraphics[height=1.5in, width=1.5in]{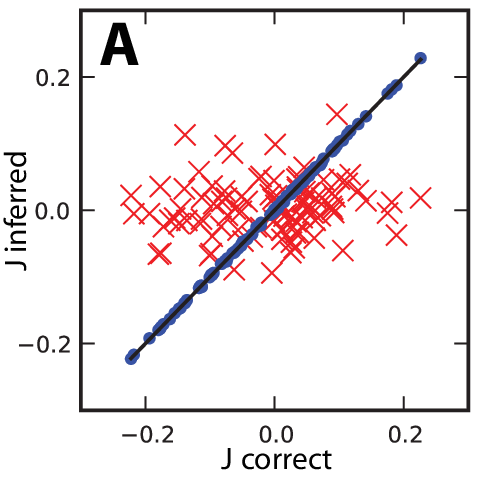}}
\subfigure{\includegraphics[height=1.5in, width=1.5in]{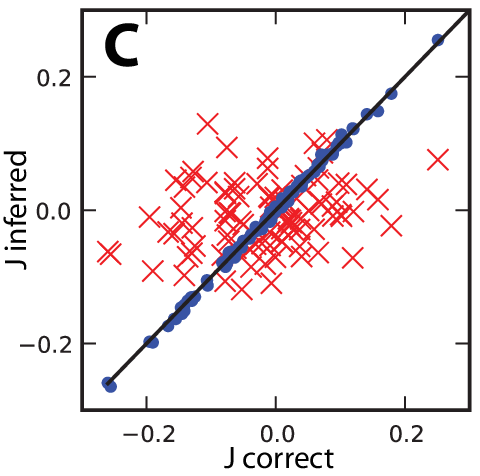}}\\
\subfigure{\includegraphics[height=1.5in, width=1.5in]{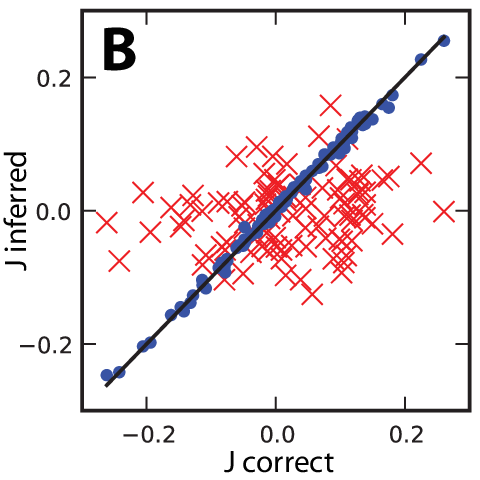}}
\subfigure{\includegraphics[height=1.5in, width=1.5in]{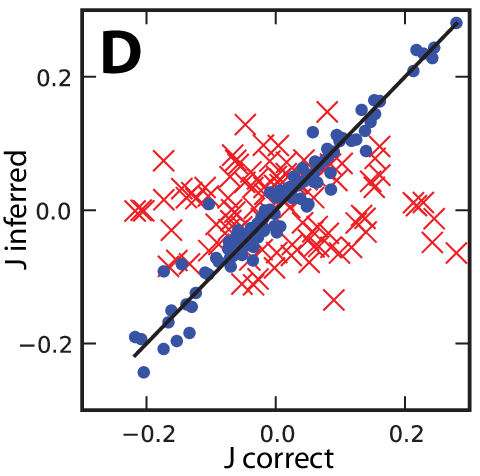}}
\caption{(Color online) Reconstructing the full network. Scatter plots showing the inferred couplings (blue dots)
versus the correct ones in a network of
$N=100$ total spins, with $10$ hidden, $J_1=1$ and data length of $T=10^6$. Red crosses 
show the initial values of the couplings used for learning. A, B, C, and D
show observed-to-observed, observed-to-hidden, hidden-to-observed and hidden-to-hidden couplings, respectively.}
\label{Fig2}
\end{figure}
The effect of TAP corrections becomes very important in reconstructing the couplings. We
found that the saddle point learning was always unstable leading to divergent couplings.
However, including the TAP corrections dramatically changed this. Fig.\ \ref{Fig2} shows
an example of this when $10\%$ of the spins were hidden. In this case, good 
reconstruction of all couplings could be achieved.

\begin{figure}[t]
\subfigure{\includegraphics[height=1.36in, width=1.62in]{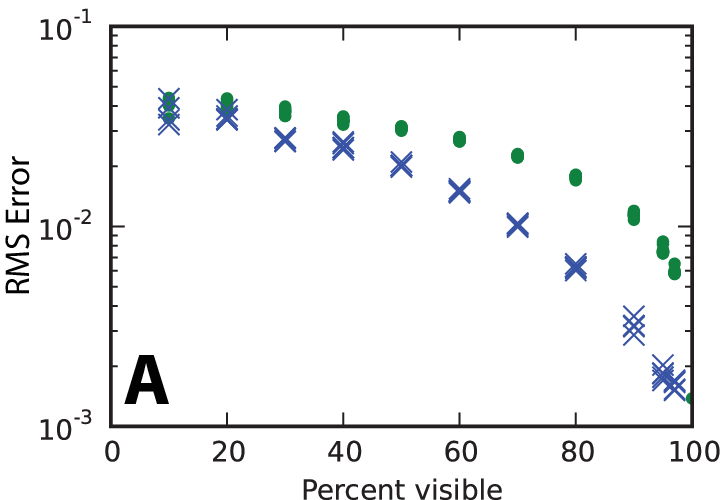}}
\subfigure{\includegraphics[height=1.36in, width=1.62in]{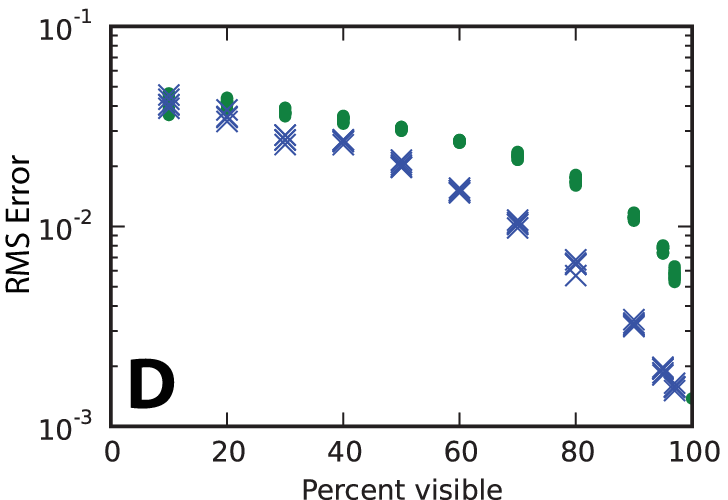}}\\
\subfigure{\includegraphics[height=1.36in, width=1.62in]{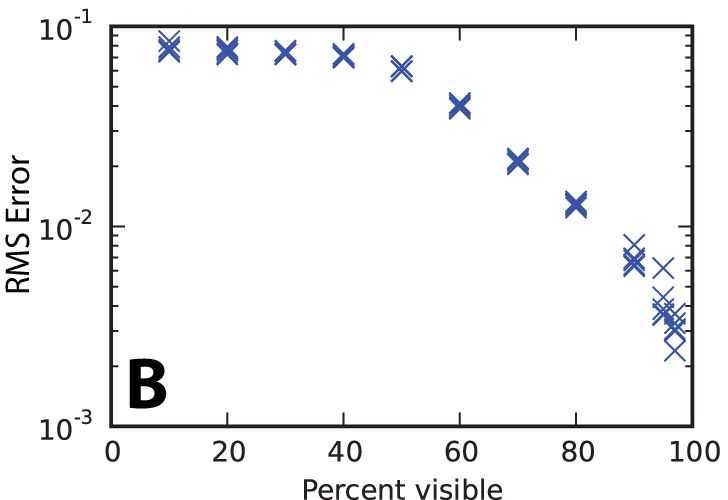}}
\subfigure{\includegraphics[height=1.36in, width=1.62in]{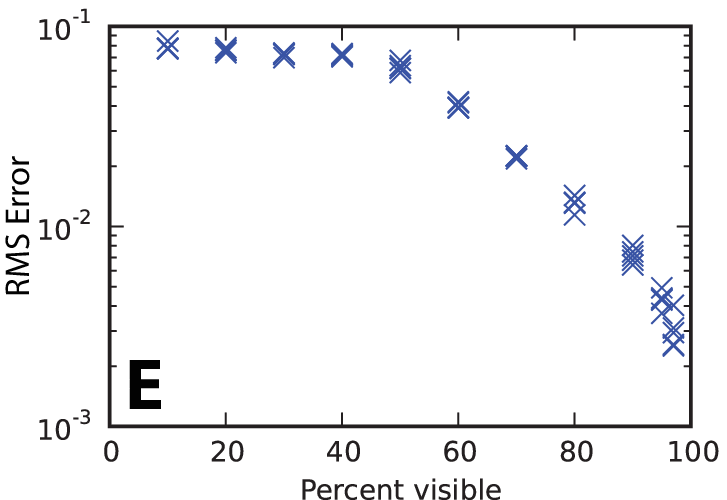}}\\
\subfigure{\includegraphics[height=1.36in, width=1.62in]{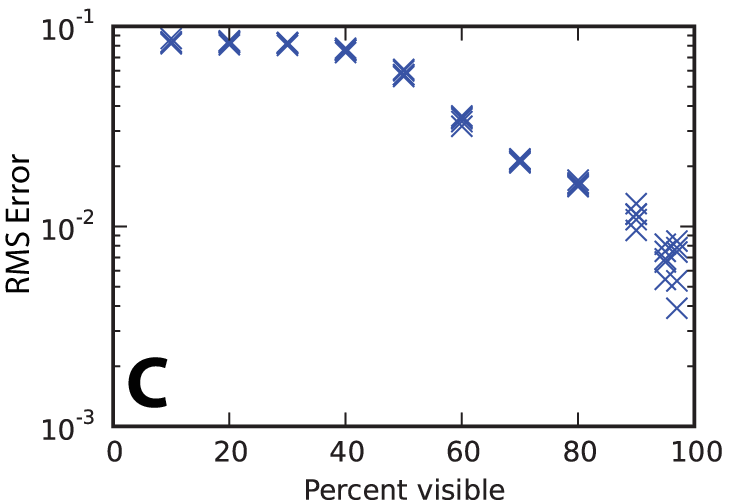}}
\subfigure{\includegraphics[height=1.36in, width=1.62in]{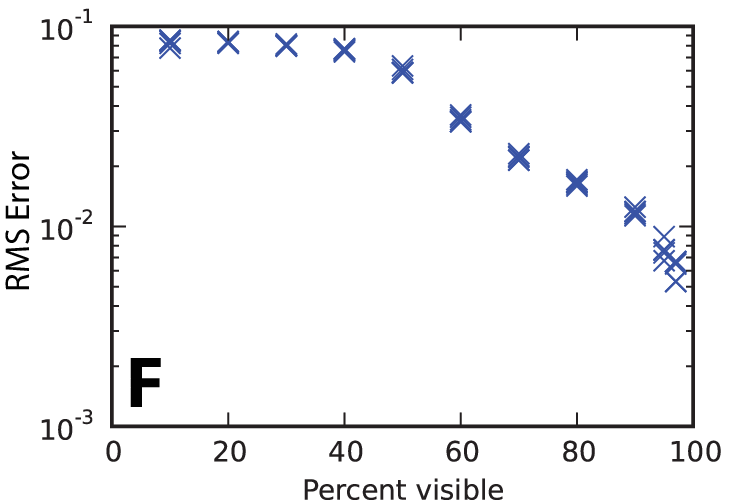}}
\caption{(Color online) Root mean squared error of the couplings. (A)-(C) show the rms error
for the observed-to-observed, hidden-to-observed and observed-to-hidden couplings using TAP learning
when the original model did not have hidden-to-hidden connections; the other couplings were drawn independently
from a Gaussian. (D)-(F) show the same but when the hidden-to-hidden connections were present in the original network but were 
ignored (i.e. put to zero) when learning the other couplings. In (A) and (D) the green dots show the inferred
couplings by maximizing the likelihood with hidden nodes completely ignored. 
Here we used $J_1=1, N=100, T=10^6$, and
five different random realizations of the connectivity.}
\label{Fig3}
\end{figure}

Although for a proportionally small number of hidden nodes we observed good reconstruction, when the hidden
part was large, for a constant data length, we found that the algorithm starts to become
unstable: the root mean squared error for the inferred hidden-to-hidden connections
increases after a few iterations and the others follow. This instability, however, seems
to arise from our attempt to learn the hidden-to-hidden connections. For an architecture
where there is no hidden-to-hidden couplings, that is a dynamical semi-restricted
Boltzmann machine \cite{Osindero08,*taylor2009factored}, or when hidden-to-hidden
couplings exist in the original network but are ignored during learning (i.e. put to zero), the algorithm
allows recovering the other connections even for very large hidden sizes. This is shown in
Fig.\ \ref{Fig3}.

\begin{figure}[t]
\centering
\subfigure{\includegraphics[height=1.5in, width=2.2in]{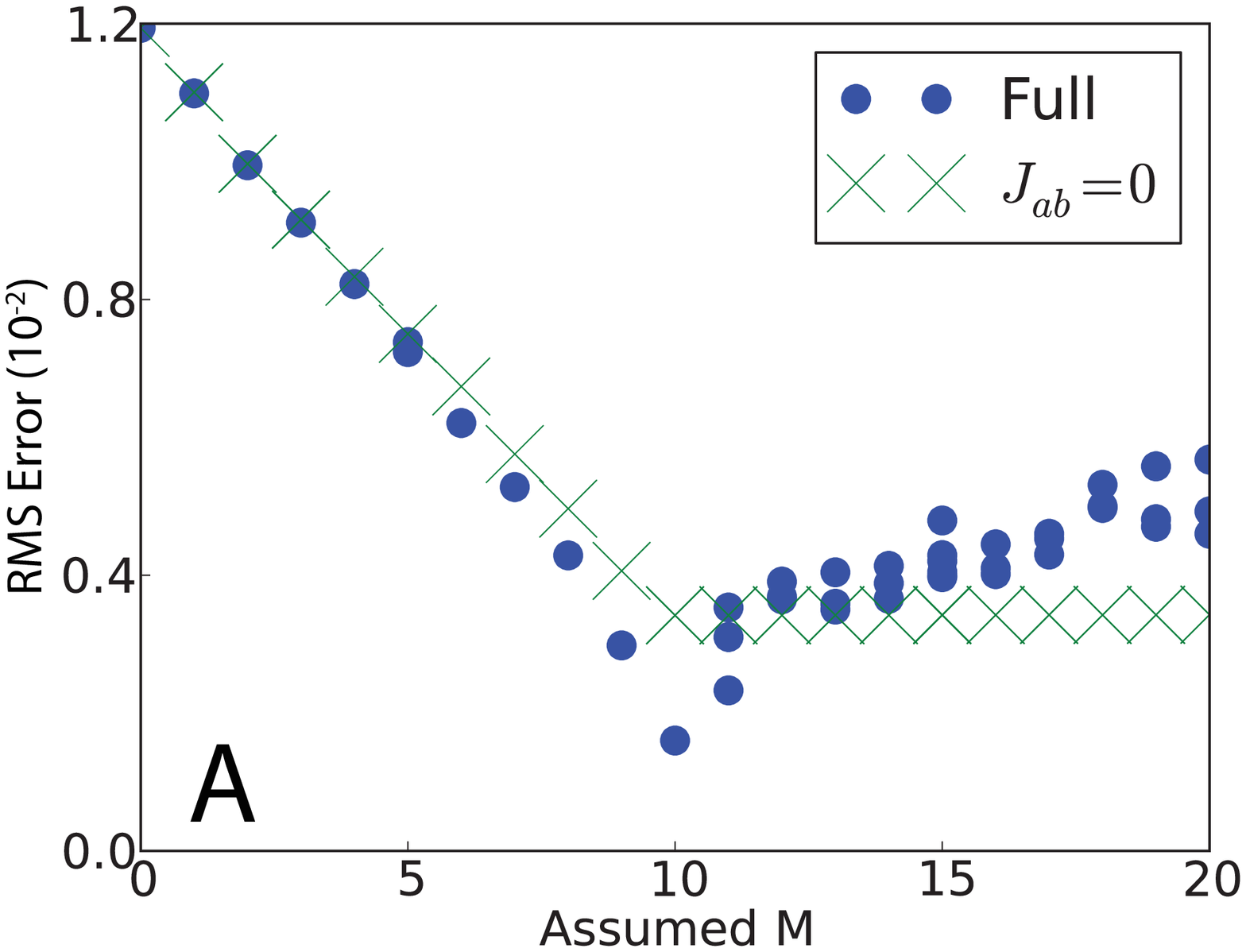}} 
\subfigure{\includegraphics[height=1.5in, width=2.2in]{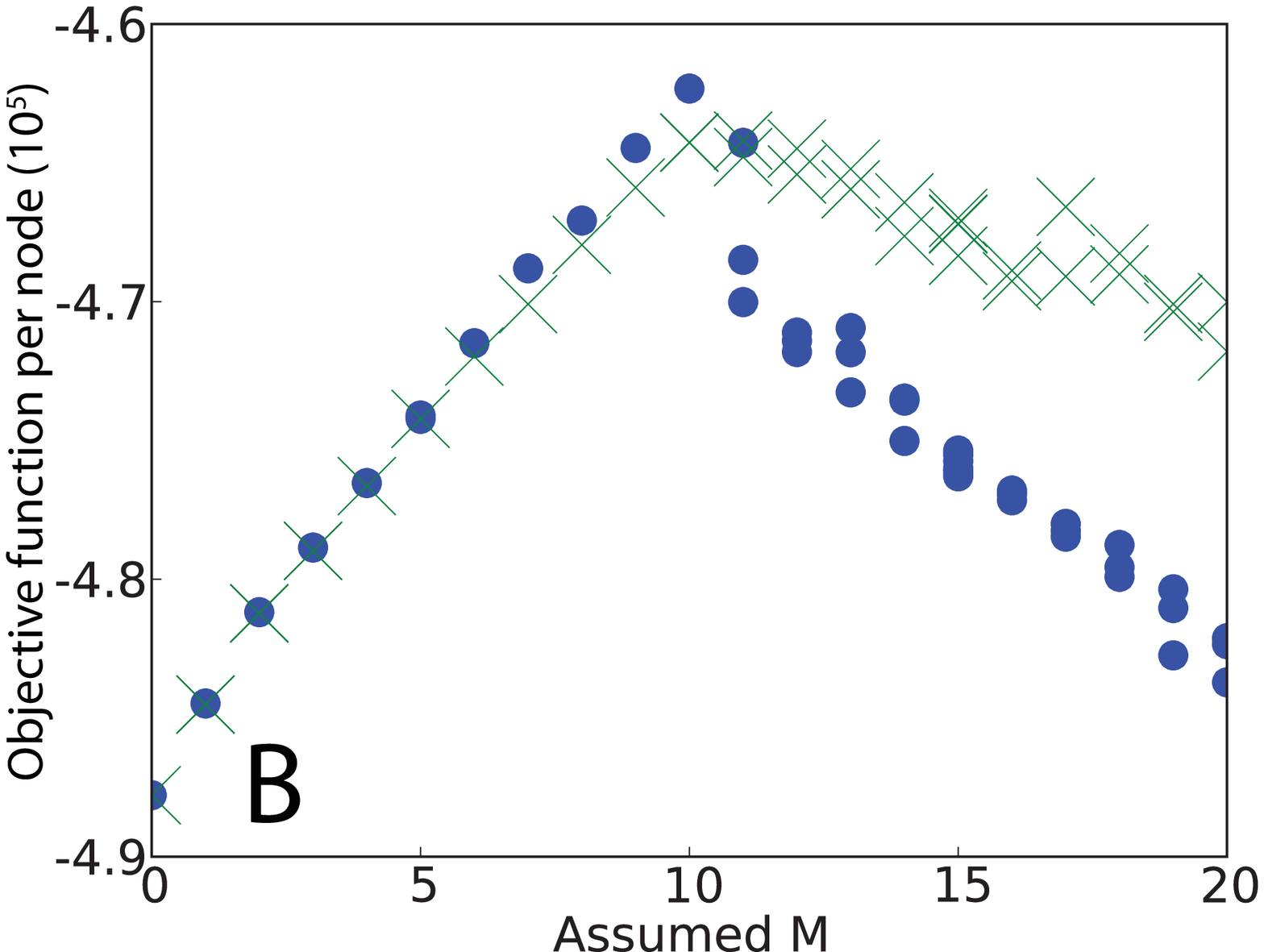}} 
\caption{(Color online) Estimation of the number of hidden nodes. 
(A) the RMS error of the observed-to-observed couplings inferred using TAP learning
for different values of assumed hidden units, $M$. The couplings are inferred either
assuming no hidden-to-hidden connectivity (green x) or by trying to learn all couplings (blue circles).
(B) the value of the objective function per inferred node for different numbers of assumed hidden units.
The data used here is the same as that of Fig.\ \ref{Fig2}, that is in reality there were $10$ hidden nodes
which is where the normalized objective function shows maximum and the RMS error shows a minimum.}
\label{FigEstM}
\end{figure}

In the examples above, we have used the right number of hidden nodes when learning the couplings. In practice, however,
the number of hidden nodes may be unknown. We thus wondered if the number of hidden nodes 
can be estimated from the data. In Fig.\ \ref{FigEstM}, we plot the objective function 
$\Gamma_1/N$ for the data used in Fig. \ref{Fig2}, where $N$ is the
total number of visible nodes, that is, the sum of the observed nodes, and an unknown number $M$
of hidden nodes. As can be seen in Fig.\ \ref{FigEstM}, varying $M$, at the correct number of hidden units
a peak in $\Gamma_1/N$ is seen and also the reconstruction error of the visible-to-visible connectivity
reaches a minimum. This was the case, even when we constrained the hidden-to-hidden connections to zero during learning.
However, the peak in the normalized cost function was sharper, and the minmum achieved for the
reconstruction lower, when hidden-to-hidden connections were also learned. 

\section{Discussions}
In this paper we described an approach for learning and inference in
a nonequilibrium Ising model when it is partially observed, that is, when the history of
spin configurations are known only for some spins and not for the others. Treating
the log-likelihood of the data as a path integral over
the fields acting on hidden and observed spins in a non-equilibrium Ising model, we
approximated it first using saddle point approximations, and then by considering TAP
corrections to it.

In usual statistical mechanics settings, TAP equations can be derived in several ways:
using Plefka expansion \cite{Plefka82,*Biroli99,*Roudi11-2}, information-geometric
arguments \cite{Kappen00} or by calculating the Gaussian fluctuations around the saddle
point \cite{Kholodenko1990,negele1998quantum}. Here we took an approach similar to the latter case
except that we did not use the Hubbard-Stratonovich transformation. It would be quite
interesting to see what the other approaches yield as TAP corrections. 
Tyrcha and Hertz \cite{TyrchaHertz12} have also studied the problem of hidden spins, 
using a different approach. Their equations agree with ours at the saddle 
point level but differ at the TAP level.

In terms of performance, saddle point equations gave results that were similar to TAP for
inferring the state of hidden spins, but were very unstable for learning the couplings.
Using TAP equations, learning all the couplings, even the hidden-to-hidden ones, was
possible when the hidden part was proportionally small. By ignoring
hidden-to-hidden couplings, even when they were present in the network, we could achieve
good performance in recovering all the other couplings, even with half
the network hidden. This also improved recovering the observed-to-observed couplings
compared to when hidden spins were ignored all together.  Our numerical results also 
indicate that even the number of hidden units can be estimated by optimization of 
the objective function. This is of particular practical revelance as often 
the true size of the network is unknown. In Fig.\ \ref{FigEstM}, both the objective function
and the reconstruction quality increase as we add hidden nodes
suggesting that even a wrong number of hidden nodes can improve the reconstruction of
observed-to-observed couplings. This can be explained by the fact that the added hidden
variables, even if wrong in number, to some degree can explain the correlations induced by
the hidden spins in the real network.

An important future direction, in our view, is to study how averaging the likelihood over
some couplings, in particular the hard to infer hidden-to-hidden ones, given a prior over
them, will influence the reconstruction of the other couplings. There is a large body of
work where the path integral approach is used to study the disorder-averaged behavior of
disordered systems \cite{de1978dynamics,*Sompolinsky81,Coolen2001619}. The approach that we
developed here can be easily combined with this work to average out some of the couplings.

\section{Acknowledgement} We are most grateful to Joanna Tyrcha and John Hertz 
for constructive comments and for discussing their work with us,
and Claudia Battistin for comments on the paper. Benjamin Dunn acknowledges the hospitality of 
NORDITA. We also acknowledge the computing resources provided by the University of Oslo
and the Norwegian metacenter for High Performance Computing (NOTUR), Abel cluster.

\appendix
\renewcommand{\theequation}{I-\arabic{equation}}
\setcounter{equation}{0}
\section{Legendre transform}
\label{appendixI}
Here we derive Eq.\ \ref{Gamma0} in the main text. Starting from Eq.\ 7, we have
\begin{widetext}
\begin{subequations}
\begin{eqnarray}
&& \calL[\psi]= \log  \int D\calG \exp[\Phi],  \\
&&\Phi=\sum_{i,t} s_i(t+1)g_i(t)+i\sum_{i,t} \hat{g}_i(t) \Big[g_i(t)-\sum_j J_{ij} s_j(t)\Big]+ i\sum_{a,t} \hat{g}_a(t)\Big[g_a(t)-\sum_i J_{ai} s_i(t)\Big] -\sum_{i,t} \log 2\cosh[g_i(t)] \nonumber \\
&&\hspace{2cm} -\sum_{a,t}\log 2\cosh[g_a(t)] + \sum_{a,t} \log 2\cosh \Big[g_a(t-1)-i\sum_b J_{ba} \hat{g}_b(t)-i\sum_i J_{ia} \hat{g}_i(t)+\psi_a(t)\Big] \label{Phiapp}\\
&&\mu_a(t)=\frac{\partial \Phi}{\partial \psi_a(t)}=\tanh\big[g_a(t-1)-i\sum_j \hat{g}_j(t) J_{ja}-i\sum_b \hat{g}_b(t) J_{ba} + \psi_a(t)\big], \label{supermu}
\end{eqnarray}
\end{subequations}
\end{widetext}
with the saddle point being at Eq.\ \ref{gsaddles}, with $\mu_a$ instead of $m_a$. Applying the following identity to the last sum in Eq.\ \ref{Phiapp} 
\begin{eqnarray}
&&\log 2 \cosh[x]=S[\tanh[x]]+x\tanh[x]\\
&&S[x]\equiv -\frac{1+x}{2}\log \left [\frac{1+x}{2}\right]-\frac{1-x}{2}\log \left [\frac{1-x}{2}\right]
\end{eqnarray}
and using Eq.\ \ref{supermu} and the saddle point values for $g_i$ and $g_a$, we obtain
\begin{eqnarray}
&&\Phi_0=\sum_{i,t} [s_i(t+1) g_i(t)- \log 2\cosh[g_i(t)]] \nonumber \\
&&\hspace{1cm} +\sum_{a,t}[\mu_a(t+1) g_a(t)-\log 2\cosh[g_a(t)]] \nonumber \\
&&\hspace{1cm} +\sum_{a,t} \mu_a(t)\psi_a(t)+\sum_{a,t} S[\mu_a(t)]
\end{eqnarray}
At the saddle point we have $\calL_0=\Phi_0$ and using $\Gamma_0=\calL_0-\sum_{a,t} \psi_a(t)\mu_a(t)$ we arrive at Eq.\ \ref{Gamma0}. 
\renewcommand{\theequation}{II-\arabic{equation}}
\setcounter{equation}{0}
\section{Corrections to the saddle-point likelihood}
\label{appendixII}
To include the Gaussian corrections in Eq.\ \ref{GammaTAP} we first 
calculate the Gaussian fluctuations around the saddle point value of $\calL_0.$
Given Eq.\ \ref{Phiapp}, we have
\begin{subequations}
\begin{eqnarray}
&&A_{ij}^{tt'}\equiv \frac{\partial^2 \Phi}{\partial  g_i(t) \partial  g_j(t')}=-\delta_{ij}\delta_{tt'} [1-\tanh^2[g_i(t)]]\nonumber \\
&&B_{ij}^{tt'}\equiv\frac{\partial^2 \Phi}{\partial  \hat{g}_i(t) \partial  \hat{g}_j(t')}=-\sum \nolimits_{a} J_{ia} J_{ja}[1-\mu^2_a(t)] \delta_{tt'}\nonumber \\
&&C_{ab}^{tt'}\equiv\frac{\partial^2 \Phi}{\partial  g_a(t) \partial  g_b(t')}\nonumber\\
&&\hspace{2cm}=-\delta_{ab}\delta_{tt'}\Big[\mu^2_a(t+1)-\tanh^2[g_a(t)]]\Big] \nonumber \\
&&D_{ab}^{tt'}\equiv\frac{\partial^2 \Phi}{\partial  \hat{g}_a(t) \partial  \hat{g}_b(t')}=-\sum \nolimits_{c} J_{ac} J_{bc} [1-\mu^2_c(t)] \delta_{tt'}\nonumber  \\
&&E_{ib}^{tt'}\equiv\frac{\partial^2 \Phi}{\partial  \hat{g}_i(t) \partial  \hat{g}_b(t')}=-\sum \nolimits_{a} J_{ia} J_{ba} [1-\mu^2_a(t)] \delta_{t,t'}\nonumber \\
&&F_{ib}^{tt'}=\frac{\partial^2 \Phi}{\partial  \hat{g}_i(t) \partial  g_b(t')}= -i J_{ib} [1-\mu^2_b(t)] \delta_{t-1,t'}\nonumber \\ 
&&\frac{\partial^2 \Phi}{\partial  g_a(t) \partial  \hat{g}_b(t')}=i\delta_{ab}\delta_{tt'} \underbrace{-iJ_{ba}[1-\mu^2_a(t+1)] \delta_{t',t+1}}_{G_{ab}}\nonumber
\end{eqnarray}
\end{subequations}
and $\partial^2 \Phi/\partial  g_i(t) \partial  \hat{g}_j(t')=i\delta_{ij}\delta_{tt'}
$leading to
\begin{equation}
  \left[
  \begin{BMAT}{cc.cc|cc.cc}{cc.cc|cc.cc}
      A^{tt}   &  i\ID &  0   &  0   & 0   & 0   & 0    & 0\\ 
      i\ID &  B^{tt}   &  0   &  E^{tt}   & 0   & 0   & 0    & 0\\ 
      0   &  0   &  C^{tt}  &  i\ID & 0   & [F^{t't}]^T   & 0    & G^{tt'}\\
      0   &  [E^{tt}]^{T}   &  i\ID &  D^{tt}   & 0   & 0   & 0    & 0\\
      0   &  0   &  0   &  0   & A^{t't'} & i\ID & 0    & 0\\
      0   &  0   &  F^{t't}   & 0   & i\ID & B^{t't'} & 0    & E^{t't'}\\
      0   &  0   &  0   &  0   & 0   & 0   & C^{t't'}    & i\ID\\
      0   &  0   &  [G^{tt'}]^{T}   &  0   & 0   & [E^{t't'}]^T  & i\ID  & D^{t't'}\\
   \end{BMAT}
 \right]
\end{equation}
as part of the Hessian for $t$ and $t'=t+1$, and the same structure gets repeated for other times.
The first to the fourth rows/columns correspond to derivatives with 
respect to $g_i,\hat{g}_i,g_a,\hat{g}_a$ all at $t$ and the fourth to eighth rows/columns 
at $t'=t+1$.

Denoting the matrix of the blocks on the diagonal part of the Hessian as $\alpha$ and the rest as
$\beta$, we write
\begin{eqnarray}
&&\log \det (\alpha+\beta)= \log \det(\alpha) + \log \det[I+\alpha^{-1}\beta]\nonumber \\
&&=\log \det(\alpha)+\Trace \log [I+\alpha^{-1}\beta]\\
&&\approx \log \det(\alpha)+\Trace [\alpha^{-1}\beta]+\frac{1}{2}\Trace \{ [\alpha^{-1}\beta]^2\} \}+ \cdots \nonumber
\end{eqnarray}
Assuming that the couplings are random and independent with mean of order $1/N$ and a
standard deviation of $J_1/\sqrt{N}$ as typically assumed in mean-field models of 
spin glasses \cite{FischerHertz},
$\log \det(\alpha)$ will be of quadratic order in $J_1$. 
Since $\alpha$ is block diagonal, so will be $\alpha^{-1}$ and therefore $\Trace [\alpha^{-1}\beta]=0$. 
Ignoring the second trace as contributing higher order terms, we thus 
approximate the determinant of the Hessian matrix by considering only the block diagonal terms 
\begin{equation}
\det(\alpha)=\prod_t \det\{A(t)B(t)+\ID\} \det\{C(t)D(t)+\ID\}
\end{equation}
\newpage
leading to the following estimate of the contribution of the Gaussian fluctuations
\begin{eqnarray}
  &&  \delta \calL\approx-\frac{1}{2}\sum_t \log \det\{A(t)B(t)+\ID\}   \label{deltaL}\\
  && \hspace{2mm}-\frac{1}{2} \log \det\{C(t)D(t)+\ID\} \nonumber \\
  &&\hspace{2mm}\approx-\frac{1}{2}\sum_{t,i} \big\{[1-\tanh^2[g_i(t)]] \sum_b J^2_{ib} [1-\mu^2_b(t)]\big\} \nonumber \\
  &&  \hspace{2mm}-\frac{1}{2} \sum_{t,a} \big\{[\mu^2_a(t+1)-\tanh^2(g_a(t))] \sum_b J^2_{ab} [1-\mu^2_b(t)]\big\}\nonumber
\end{eqnarray}
With these Gaussian fluctuations taken into account, the corrected value for the mean magnetization is
\begin{eqnarray}
&&m_a(t)=\frac{\partial \calL}{\partial \psi_a(t)}=\mu_i(t)+l_a(t)\\
&&l_a(t)=\frac{\partial \delta \calL}{\partial \psi_a(t)}
\end{eqnarray}
We now express $\Gamma_0$ to the quadratic order in $J$ in terms of $m_a$ instead of $\mu_a$. With some 
algebra at $\psi \to 0$ we have
\begin{eqnarray}
&&\Gamma_0[m]=\sum_{i,t} [s_i(t+1) g_i(t)-\log 2\cosh(g_i(t))]\nonumber \\
&&+\sum_{a,t} [m_a(t+1) g_a(t)-\log 2\cosh(g_a(t))]+\sum_{a,t} S[m_a(t)]\nonumber\\
&&-\sum_{i,t,a} [s_i(t+1) -\tanh(g_i(t))] J_{ia} l_a(t)\nonumber\\
&&+\sum_{a,t} l_a(t+1) [\tanh^{-1}[m_a(t+1)]-g_a(t)]\nonumber\\
&&-\sum_{a,t,b} [m_a(t+1) - \tanh[g_a(t)] J_{ab} l_b(t)\nonumber
\end{eqnarray}
Where we have abused the notation here, using $g_a$ and $g_i$ to indicate the fields calculated at 
$m_a(t)$ instead of $\mu_a(t)$. Noting that $l_a$ itself is quadratic in $J$, the terms that involve $l_a$ above
are all higher order, and can therefore be ignored. This, combined with the expression
for $\delta \calL$ in Eq.\ \ref{deltaL} yield Eq.\ \ref{GammaTAP} in the main text.
\bibliography{mybibliography2010}
\end{document}